\documentstyle[aaspp4,emulateapj5,apjfonts]{article}

\gdef\sfr{$M_{\odot}$\,yr$^{-1}$}
\gdef\h50min{$h_{50}^{-1}$}

\gdef\3727{[O\,{\sc ii}]\,3727\,\AA}
\gdef\5007{[O\,{\sc iii}]\,5007\,\AA}

\gdef\ms1054{MS\,1054--03}
\gdef\4ang{4000\,\AA}
\gdef\mperyr{$M_{\odot}$\,yr$^{-1}$}
\gdef\fluxunit{ergs\,s$^{-1}$\,cm$^{-2}$}
\gdef\lumunit{ergs\,s$^{-1}$}
\lefthead{Rubin et al.}
\righthead{Chandra Constraints on Red Galaxies at $z\gtrsim 2$}
\slugcomment{Accepted for publication in the Astrophysical Journal Letters}
\begin{document}

\title{Chandra Constraints on the AGN
Fraction and Star Formation
Rate of Red $z\gtrsim 2$ Galaxies in the FIRES MS\,1054--03
Field}

\author{Kate H. R. Rubin\altaffilmark{1},
Pieter~G.~van Dokkum\altaffilmark{1}, 
Paolo Coppi\altaffilmark{1},
Olivia Johnson\altaffilmark{2},
Natascha M.~F\"orster Schreiber\altaffilmark{3},
Marijn Franx\altaffilmark{4},
and
Paul van der Werf\altaffilmark{4}
}

\altaffiltext{1}{Department of Astronomy, Yale
University, New Haven, CT 06520-8101}
\altaffiltext{2}{Institute for Astronomy, University of Edinburgh,
Edinburgh EH9 3HJ}
\altaffiltext{3}{Max-Planck-Institut f\"ur extraterrestrische Physik,
Postfach 1312, 85741 Garching, Germany}
\altaffiltext{4}{Leiden Observatory, P.O. Box 9513, NL-2300 RA, Leiden,
The Netherlands}

\begin{abstract}

Very deep near-infrared observations in the
Faint InfraRed Extragalactic Survey (FIRES) have recently uncovered
a significant population of red galaxies at redshifts $z>2$. These
distant red galaxies (DRGs) are efficiently selected by the
criterion $J_s-K_s>2.3$.
We use Chandra data to examine the X-ray emission from DRGs
in the $5'
\times 5'$ FIRES MS\,1054--03 field.
Two of 40 DRGs with $K_s<22$ are detected by Chandra, and we infer
that $5^{+3}_{-2}$\,\% of DRGs host active nuclei with
$L_X>1.2 \times 10^{43}$\,\lumunit. This fraction is smaller than
that inferred from optical and near-IR spectroscopy, probably largely
due to strong spectroscopic selection biases. 
By stacking all undetected DRGs we find that their average X-ray
flux in the $0.5-8$\,keV band is $\approx 4.6 \times 10^{-17}$\,\fluxunit.
The detection is only significant
in the soft ($0.5-2$\,keV; $3.4\sigma$) and full ($0.5-8$\,keV;
$3.2\sigma$) energy bands.
The mean detection may result from star formation, the 
presence of low luminosity AGN, or a combination of both.
Assuming the detection is due exclusively to star formation,
we find an average instantaneous star formation rate of
$214 \pm 68$ (random) $\pm 73$ (systematic)
\sfr, in excellent agreement with
previous results from spectral energy distribution
fitting when constant star formation
histories are assumed. These results may imply that
DRGs contribute significantly to the cosmic star formation
rate at $z\approx 2.5$. However, the mean X-ray flux
strictly provides only an upper limit to the star formation
rate due to the uncertain contribution
of low luminosity, possibly obscured AGNs.
Observations at other wavelengths are needed to provide 
independent estimates
of the star formation rate of DRGs.

\end{abstract}

\keywords{cosmology: observations ---
galaxies: evolution --- galaxies:
formation
}


\section{Introduction}

Recent studies have demonstrated that galaxies at $z>2$ can be
efficiently selected by the simple observed near-infrared (NIR) color
criterion $J_s-K_s>2.3$ ({Franx} {et~al.} 2003; {van Dokkum} {et~al.}
2003). These distant red galaxies (DRGs) are typically very faint in
the rest-frame ultra-violet (UV), and are complementary to the
UV-selected Lyman break galaxies (LBGs). Studies of their broad-band
spectral energy distributions (SEDs) (F\"orster Schreiber et al.\ 2004a) and
their rest-frame optical emission lines (van Dokkum et al.\ 2004)
indicate that at a given rest-frame optical luminosity DRGs are
dustier, more massive, and have higher ages than LBGs.
Their star formation rates are quite uncertain,
as they depend on
the assumed star formation histories: median values for DRGs
with $K<21.7$ range from $\sim
170$\,\mperyr\ for constant star formation histories to $\sim
23$\,\mperyr\ for declining models (F\"orster Schreiber et al.\
2004a).

In this paper we study the X-ray emission from DRGs, using Chandra
observations of the field centered on the foreground cluster
MS\,1054--03 at $z=0.83$. This field is one of the two
Faint InfraRed Extragalactic Survey (FIRES) fields and its population
of red $z>2$ galaxies has been described by van Dokkum et al.\ (2003; 2004)
and F\"orster Schreiber et al.\ (2004a). The X-ray properties provide
information on the fraction of Active Galactic Nuclei (AGN) among
DRGs. Spectroscopy indicates that this fraction could be substantial
(van Dokkum et al.\ 2003, 2004), but this interpretation is uncertain
because current spectroscopic samples are very incomplete.
Furthermore, the stacked X-ray flux of undetected sources provides
an additional constraint on the average instantaneous star formation rate.
Where needed we assume
$\Omega_m=0.3$, $\Omega_{\Lambda}=0.7$, and $H_0=70$\,km\,s$^{-1}$\,Mpc$^{-1}$.
Galaxy identifications refer to the F\"orster Schreiber et al.\ (2004b)
FIRES catalog of the MS\,1054--03 field, which includes photometric
redshifts for all sources.\vspace{-0.2cm}

\section{Observations and Sample Selection}

The X-ray data were obtained from an archival 91\,ks exposure with the 
Chandra ACIS-S3 detector. The data were obtained by the ACIS GTO
team to study the foreground cluster (Jeltema et al.\ 2001).
The reduction of the data and an analysis of the point sources in
the field are presented in Johnson, Best, \& Almaini (2003).
After removal of periods with strong background flaring the
effective exposure time is 74\,ks.
Images of the full energy band (0.5--8 keV) were
made, as well as of the soft (0.5--2 keV) and hard (2--8 keV) bands.

The DRGs were selected from the $5' \times 5'$ area imaged with
the VLT Infrared Spectrograph 
And Array Camera in the $J_s$, $H$ and $K_s$ bands as part of
the FIRES project.
HST WFPC2 data is available
in the $V_{606}$ and $I_{814}$ bands, as well as VLT FORS1
data in the Bessel $U$, $B$ and $V$ bands.
The reduction, source detection, and photometry are described
in F\"orster Schreiber et al.\ (2004b). There are 45 sources
with $J_s-K_s \ge 2.3$, $K_s \le 22$, and photometric weights
(i.e., the fraction of the maximal exposure time spent on-source)
in $J_s$ and $K_s$ greater than $0.1$.
Figure \ref{jk.plot} shows the selection limits.
From this group, we removed
2 sources because they have photometric redshifts $z<1.8$, and
three sources within $\sim 1'$\
of the center of the cluster,
to minimize effects of the diffuse cluster light.
The final sample thus consists of 40 DRGs.

\vbox{
\begin{center}
\leavevmode
\hbox{%
\epsfxsize=8cm
\epsffile{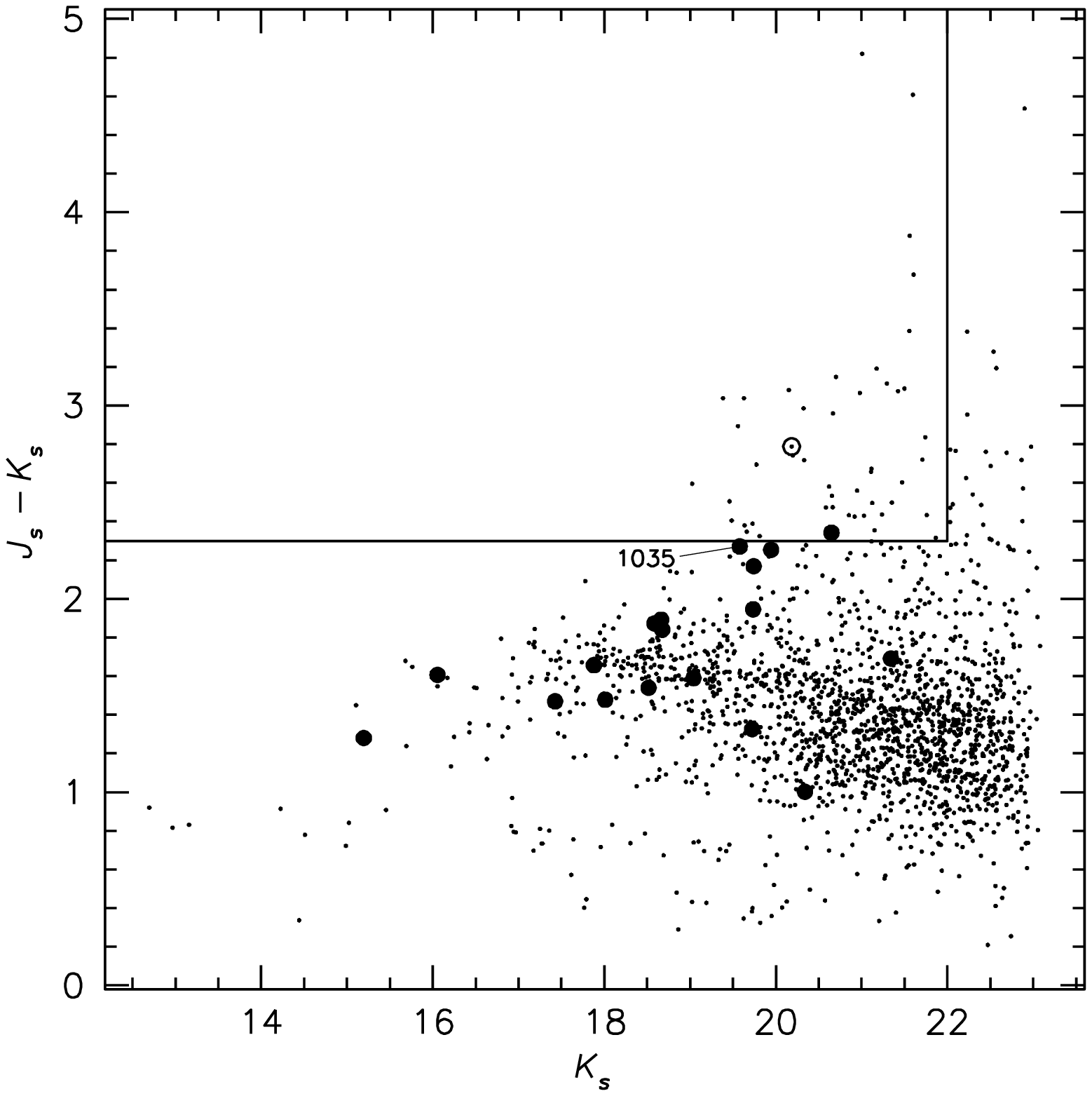}}
\figcaption{\small
Color-magnitude diagram for all objects in the VLT field.  Chandra point 
sources from Johnson et al.\ (2003) are shown as solid circles.  The open 
circle shows object 313, which is a $4.7\sigma$ detection.  Our selection
criteria are shown with lines.
The fraction of Chandra sources among galaxies with $J_s-K_s>2.3$ and
$K_s<22$ is $5$\,\%.
\label{jk.plot}}
\end{center}}

\section{Direct Detections: AGN Fraction}

Johnson et al.\ (2003) selected point-sources in the MS\,1054--03 field
using WAVDETECT with a rather conservative threshold (giving
$\sim 1$ spurious source over the entire field),
and we first determined the overlap between DRGs at $K_s<22$ and
X-ray sources in the Johnson et al.\ catalog
(large solid circles in Fig.\ \ref{jk.plot}).
Only one DRG (object 1100; $z_{\rm phot}=2.9$; $K_s=20.7$)
was found to contain an X-ray source identified
by Johnson et al.\ (object 30 in their catalog). 

Next we examined the prevalence of X-ray sources fainter than the
conservative limits imposed by Johnson et al. The X-ray
fluxes were determined using apertures of $4\arcsec$ diameter, which
contain $\gtrsim 70$\,\% of the flux for point-sources
over our entire $5'\times 5'$ FIRES
field 
\footnote{http://cxc.harvard.edu/cal/Acis/Cal\_{}prods/psf/Memo/s8.html}. 
For each DRG a local background
was determined from the mean flux in 30 randomly placed
apertures near the location of the DRG. The locations of these
apertures were constrained to be non-overlapping,
at a distance of $7\arcsec - 20\arcsec$
from the DRG, and at least
$4.5\arcsec$ from point 
sources identified by Johnson et al. 
In addition to galaxy 1100 we find one
other object (313; $z_{\rm phot}=2.0$; $K_s=20.2$) with a significant
($\approx 4.7 \sigma$) X-ray flux.

The SEDs of both objects are shown in Fig.\ \ref{seds.plot},
along with the median SED of all other DRGs (in the observed frame;
for rest-frame SEDs see F\"orster Schreiber et al.\ 2004a).
The SED of object 313  is very similar to the median,
suggesting that the AGN does not contribute
significantly to the continuum flux. However, galaxy 1100 has much
bluer optical-NIR colors than the majority of DRGs
indicating that its active nucleus
may effect the SED, in particular in the rest-frame ultra-violet.

The $4\sigma$ limit of 5.3 counts (after background subtraction)
corresponds to an X-ray flux of
$3.7 \times 10^{-16}$\,\fluxunit\ (see below), and
a K-corrected luminosity of $1.2 \times 10^{43}$\,\lumunit\
in the rest-frame $2-10$\,keV band
for the median $\langle z_{\rm phot}\rangle = 2.4$
of the DRGs in our sample.
We conclude that the fraction of AGN with 
$L_X > 1.2 \times 10^{43}$\,\lumunit\
among DRGs is $5^{+3}_{-2}$\,\% (assuming Poisson statistics).

We note here that both {\em spectroscopically}-identified AGNs in the
MS\,1054--03 field have dropped out of the DRG sample after
recalibration of their photometry (see van Dokkum et al.\ 2004).
Interestingly only one of the two (object 1035, with $J_s-K_s=2.27$;
labeled in Fig.\ 1) is detected with Chandra, suggesting that a
multi-wavelength approach is necessary to accurately determine the AGN
fraction among red galaxies at $z>2$.\vspace{0.2cm}

\vbox{
\begin{center}
\leavevmode
\hbox{%
\epsfxsize=8cm
\epsffile{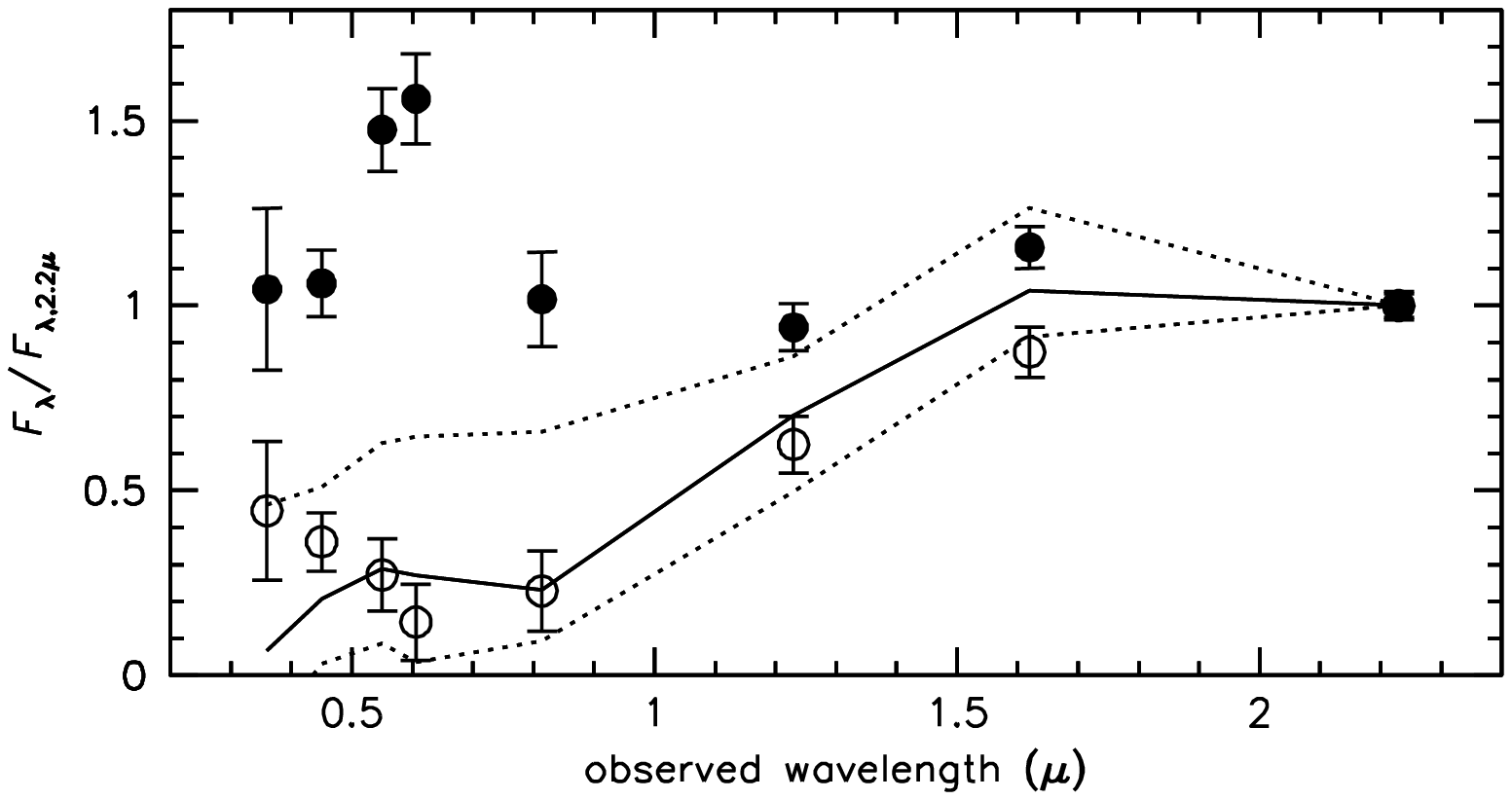}}
\figcaption{\small
Spectral energy distributions of the two DRGs associated with X-ray
sources of $L_X>1.2 \times 10^{43}$\,\lumunit. The solid line shows the
median SED of all DRGs; dotted lines show the 25- and 75-percentiles
of the distribution. The SED of 313 (open points) is very similar to those
of other DRGs, but the SED of 1100 (filled points) is likely affected
by emission
from the nucleus, in particular in the UV.\vspace{-0.3cm}
\label{seds.plot}}
\end{center}}

\section{Stacked emission of non-detections}

The low X-ray background allows very efficient stacking analyses of
sources that individually are too faint to be detected (e.g.,
Alexander et al.\ 2003; Brandt et al.\ 2001; Malhotra et al.\ 2003;
Reddy \& Steidel 2004).
We averaged the 38 flux measurements of the undetected DRGs, effectively
increasing the exposure time to 2.8\,Ms. We also averaged each of the 30
sets of background measurements. These 30 random stacks have the
same properties (background and point spread function) as the red
galaxy stack, and are used to assess the significance of the flux
in the averaged red galaxy apertures.

The results are listed in Table 1. 
The average background and significance in each
band was determined in the following way. Each of the 30
sets of random apertures was averaged in the same way as the
DRG apertures, providing 30 independent measurements of the mean
of 38 random apertures. The mean of these 30 measurements gives
the background value, and their rms is used to calculate
the significance of the mean detection.
After background subtraction, the mean emission from DRGs is
0.54 counts (3.2$\sigma$ significance) in the full band,
0.41 counts (3.4$\sigma$) in the soft band, and
0.14 counts (1$\sigma$) in the hard band.
Images of the average detections are shown in Fig.\ \ref{stack.plot}.

A concern is that the mean flux is dominated by one or two
luminous sources which
are just below the detection threshold. Figure \ref{fluxhist.plot}
shows the distribution of the X-ray fluxes of our undetected DRG
sample after background subtraction, as well as the average
distribution of the 30 random samples.  The DRGs show a tail of
positive fluctuations and a deficit of negative fluctuations,
consistent with the hypothesis that most sources contribute to
the mean detection.

\vbox{
\begin{center}
\leavevmode
\hbox{%
\epsfxsize=8cm
\epsffile{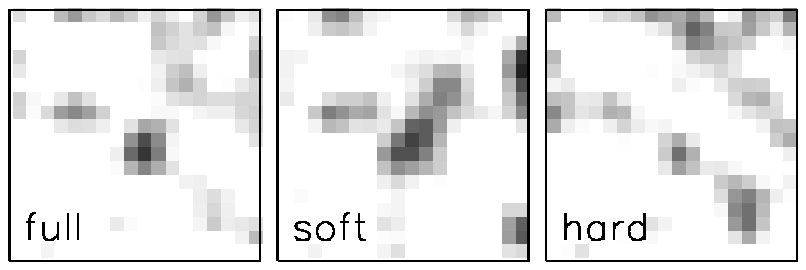}}
\figcaption{\small
Stacked images of DRGs in the full ($0.5-8$\,keV), soft ($0.5-2$\,keV)
and hard ($2-8$\,keV) bands. The size
of each image is $10'' \times 10''$. The DRGs are detected
in the soft (3.4 $\sigma$) and full bands (3.2 $\sigma$). The detection
in the hard band is not significant.
\label{stack.plot}}
\end{center}}

\vbox{
\begin{center}
\leavevmode
\hbox{%
\epsfxsize=8cm
\epsffile{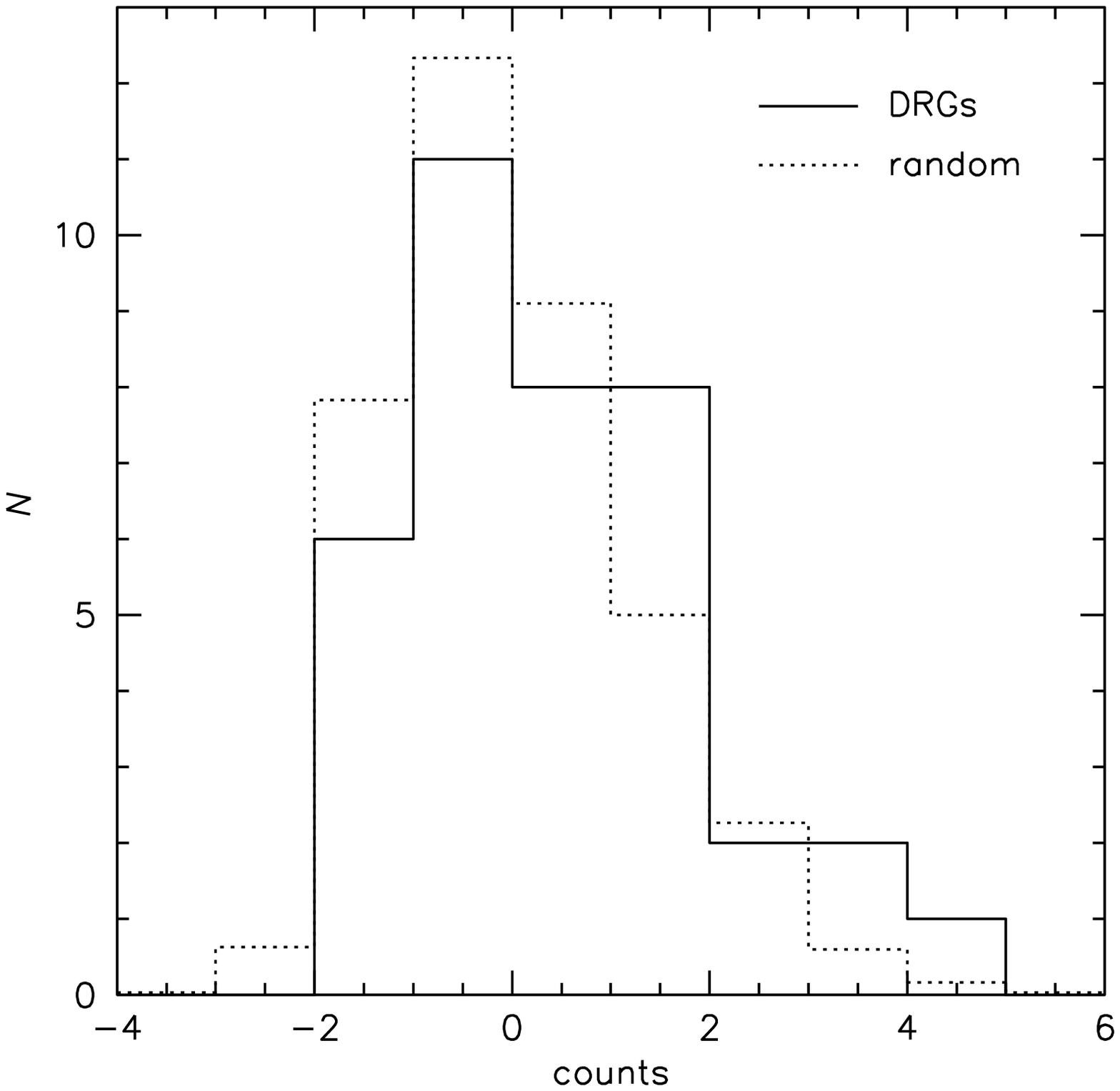}}
\figcaption{\small
Distribution of X-ray emission among red galaxies (solid line) and the average
histogram for the random samples (broken line), after background
subtraction. Compared to the random apertures
the DRGs show a deficit of negative fluctuations and
an excess of positive fluctuations.
\label{fluxhist.plot}}
\end{center}}


We empirically test how much flux is lost by our adopted $4\arcsec$
apertures by repeating our measurements of the mean X-ray emission
using differently sized apertures.  The results are listed in Table 1.
As expected, the flux is a function of aperture size. Based on
the values in Table 1 and the ACIS documentation we estimate
that the $4\arcsec$ apertures miss $30\pm 10$\,\% of the flux
at the average location of the DRGs ($\sim
2\farcm 5$ from the optical axis). This aperture effect is partially
countered by weak lensing of the background DRGs by the foreground
$z=0.83$ cluster. This effect is $\approx 15$\,\% on average
(see F\"orster Schreiber et al.\ 2004a).
The aperture- and lensing-corrected mean flux of the DRGs is 
$0.66 \pm 0.21$ (random) $\pm 0.11$ (systematic) counts
in the full band. The systematic error reflects the uncertainties
in the two corrections.


The average X-ray detection of the DRGs provides
a constraint on their instantaneous star formation rate.
In general, X-ray emission
of galaxies is due to a combination of hot gas (e.g., from
supernova bubbles and winds), low mass X-ray binaries,
high mass X-ray binaries (HMXBs), and possibly an active nucleus.
HMXBs are short-lived as they trace the prompt formation rate of
massive stars with $M_* \sim 8 M_{\odot}$, and it appears
that for galaxies with star formation
rates in excess of $\sim 5$\,\sfr\ they dominate the X-ray
luminosity, particularly above 2\,keV (Grimm et al.\ 2003).
Several recent studies have therefore explored the relation between
the instantaneous star formation rate and the rest-frame
2--10\,keV X-ray luminosity (e.g., Ranalli et al.\ 2003,
Grimm et al.\ 2003). Here we use the Grimm et al.\
(2003) relation, SFR\,$\approx 1.49 \times 10^{-40} \times
L_{2-10\,{\rm keV}}$. The Grimm et al.\ calibration
gives $\sim 30$\,\% lower star formation rates than the
Ranalli et al.\ calibration.

We convert the observed counts in the $0.5-8$\,keV energy band to
a rest-frame $2-10$\,keV luminosity in the following way. First
we determined the average conversion from counts to flux from
the sources listed in Johnson et al.\ (2003). The conversion varies
by less than 10\,\% over the ACIS field.
Their conversion assumes
a spectrum with photon index $\Gamma = 1.7$, and we use
the Portable Interactive
Multi-Mission Simulator (PIMMS) to calculate the
effect of changing $\Gamma$ from 1.7 to 2.0,
which is the appropriate value for the $>2$\,keV
spectrum of HMXBs. Applying this $\approx 15$\,\%
correction to the 
empirically determined conversion factor we
obtain $F_X (0.5-8\,{\rm keV}) = 6.97 \times {\rm counts}\,(0.5-8\,{\rm keV})$,
with $F_X$ in units of $10^{-17}$\,\fluxunit. The lensing- and
aperture-corrected mean DRG flux is then $4.6 \times 10^{-17}$\,\fluxunit\
in the 0.5--8\,keV band. Next, the flux is K-corrected to the
rest-frame 2--10\,keV band (again assuming $\Gamma=2.0$), using
$z=2.4$. Converting the rest-frame flux to a rest-frame luminosity
(for $z=2.4$) we find $L_X = 1.4 \times 10^{42}$\,\lumunit, and
the Grimm et al.\ (2003) calibration gives an instantaneous
star formation rate of $214 \pm 68$ (random) $\pm 73$ (systematic)
\sfr. The systematic error is due to
uncertainties in the aperture- and lensing correction,  uncertainties
in the conversion of  X-ray luminosity to
star formation rate, and the scatter
in the relation between X-ray luminosity and star formation rate.
It was assumed that these uncertainties can be added in quadrature.
We note that PIMMS predicts a $\sim 4:1$ ratio for the soft vs.\
hard band counts, consistent with our non-detection in the hard band.

\section{Discussion}

Our estimate of the star formation rate in DRGs based on X-ray
emission can be compared to previous estimates based on SED fitting
and emission lines (van Dokkum et al.\ 2004; F\"orster Schreiber et
al.\ 2004a).  The main uncertainty in the SED fits is the assumed star
formation history. F\"orster Schreiber et al.\ (2004a) find a median
(mean) instantaneous star formation rate of $170$ (190)\,\mperyr\ for
DRGs with $K_s<21.7$ in the MS\,1054--03 field assuming continuous
star formation models, and only 23 (69)\,\mperyr\ for declining models
with $\tau = 300$\,Myr. As shown in van Dokkum et al.\ (2004) the
inclusion of H$\alpha$ measurements does not break the degeneracy
between star formation history, dust, and instantaneous star
formation rate.  Our result is in excellent agreement with the high
star formation rates derived from continuous (dusty) models,
and  may imply that DRGs contribute significantly
to the cosmic star formation rate at $z\approx 2.5$.

It is also interesting to compare
the mean X-ray luminosity of DRGs with that of Lyman break galaxies
(LBGs).  Reddy \& Steidel (2004) find a
mean X-ray flux of $\sim 5 \times 10^{-18}$\,\fluxunit\ in the
$0.5-2$\,keV band and an implied SFR of
$47$\,\mperyr\ for
LBGs at $2<z<2.5$ in the Chandra Deep Field North.
As the redshift range of these LBGs and our DRGs
is roughly similar we can compare the
fluxes directly, limiting systematic uncertainties.
The mean flux of the DRGs is $2.3\times 10^{-17}$\,\fluxunit\
in the $0.5-2$\,keV band, a factor $\sim 5$ higher than that of
LBGs.
The interpretation is unclear, as the two samples have not been
matched in observed $K$ magnitude, their luminosity functions may
be different, and the space density of DRGs is still uncertain.

Our estimate does not include one significant systematic
uncertainty, likely to be present in most attempts to infer high redshift
star formation rates from X-ray luminosities. For nearby
galaxies the angular resolution of Chandra can be used to subtract
out the possible contamination from a central active nucleus, but this
is not possible for high redshift objects. All the
X-ray luminosity attributed to HMXBs could in fact come from a population
of low
luminosity AGN. Since we do not have optical spectra for most
of our obects and we do not have the X-ray counts and count rates
to impose constraints based on X-ray spectral and variability information,
we cannot directly rule this possibility out.
 
As a general caution, we remark that even if we did have optical
spectra, it might be quite difficult to rule out the presence
of a low luminosity AGN.  If we assume that the relation between the
mass of a galaxy bulge and the mass of the bulge's central black hole
(Ferrarese \& Merritt 2000; Gebhardt et al.\ 2000)
is already established for these objects,
then the red galaxies in our sample have
typical black hole masses $\sim 10^{8}\,M_{\odot}.$ The mean X-ray
luminosity of $\sim 10^{42}$\,\lumunit\ is a
factor $\sim 100$ below the typical X-ray luminosity for unobscured
AGN with similar black hole masses, i.e., the nuclei of these galaxies
are heavily obscured or the possible black holes in these galaxes are
accreting at low, sub-Eddington rates.  In either case, if one does
not have the angular resolution to separate out the nucleus, the
starlight from the bulge can completely dominate the AGN light (e.g.,
Moran et al. 2002). The lack of obvious AGN features in a spectrum
is thus
not conclusive evidence against an AGN contribution if the aperture
used contains most of the galaxy and the signal to noise ratio of the
spectrum is low.  A better (but still circumstantial) argument
that the X-ray luminosity represents star formation might instead be
the concordance of several independent star formation indicators,
e.g., between the radio, de-reddened UV, and X-ray in Reddy \&
Steidel (2004). Further independent constraints on the star formation rate
in DRGs are thus urgently needed.

Studies of larger samples will provide a better
measurements of the number of {\em luminous} AGNs: whereas our
initial spectroscopy indicates AGN fractions of $30-50$\,\%
(van Dokkum et al.\ 2003, 2004) the Chandra data suggest that
this is largely due to the strong spectroscopic
selection bias toward galaxies with bright emission lines. 
It is also
important to directly assess the effect of AGNs
on the broad-band SEDs (and
hence estimates of star formation rates, ages, and dust content).
Planned observations
with NICMOS and ACS on HST of MS\,1054--03 will provide
constraints on the point-source contribution to the continuum
from the rest-frame UV to the optical.
Finally, deeper X-ray
observations are needed to further constrain the hardness ratio
of the mean detection and to detect more individual
X-ray sources; our results and those of Barger et al.\ (2003) suggest
that in 1--2\,Ms exposures many of the
$z>2$ red galaxies may be (faint) Chandra sources.

\acknowledgements{
We thank the anonymous referee for comments that improved
the clarity of the paper. Support from Chandra X-ray
Observatory Award AR4-5011X is gratefully acknowledged.
}


\begin{small}
\begin{center}
{ {\sc TABLE 1} \\
\sc Stacked X-ray Emission of DRGs} \\
\vspace{0.1cm}
\begin{tabular}{lccccc}
\hline
\hline
Band & Source & $2''$
& $3''$ & $4''$ & $6''$\\
\hline
Soft & $J-K$ & 0.36 & 0.76 & 1.13 & 2.05 \\
($0.5-2$\,keV)     & Backgr.\ & 0.18 & 0.41 & 0.72 & 1.62 \\
     & RMS & 0.07 & 0.11 & 0.12 & 0.19 \\
Hard & $J-K$ & 0.33 & 0.72 & 1.18 & 2.55 \\
($2-8$\,keV)     & Backgr.\ & 0.26 & 0.61 & 1.04 & 2.37 \\
     & RMS & 0.06 & 0.09 & 0.14 & 0.22 \\
Full & $J-K$ & 0.68 & 1.48 & 2.30 & 4.60 \\
($0.5-8$\,keV)     & Backgr.\ & 0.43 & 1.02 & 1.76 & 3.99 \\
     & RMS & 0.09 & 0.14 & 0.17 & 0.28 \\
\hline
\end{tabular}
\end{center}
\end{small}

\end{document}